\documentclass[11pt]{article}
\usepackage[latin1]{inputenc}
\usepackage[english]{babel}
\usepackage[namelimits]{amsmath}
\usepackage{authblk}
\usepackage{amssymb}
\usepackage{amsmath}
\usepackage{amsthm}

{\tiny }

\begin{document}
\title{{\bf On a physical description and origin of the cosmological constant}}
\author[1,2,3]{S. Viaggiu\thanks{viaggiu@axp.mat.uniroma2.it and s.viaggiu@unimarconi.it}}
\affil[1]{Dipartimento di Fisica Nucleare, Subnucleare e delle Radiazioni, Universit\'a degli Studi Guglielmo Marconi, Via Plinio 44, I-00193 Rome, Italy.}
\affil[2]{Dipartimento di Matematica, Universit\`a di Roma ``Tor Vergata'', Via della Ricerca Scientifica, 1, I-00133 Roma, Italy.}
\affil[3]{INFN, Sezione di Napoli, Complesso Universitario di Monte S. Angelo,
	Via Cintia Edificio 6, 80126 Napoli, Italy.}

\date{\today}\maketitle
\begin{abstract}
\noindent In this paper we use and extend the results present in \cite{1,2,3,4} and in particular in \cite{4} to obtain a statistical description of the cosmological constant in a cosmological de Sitter universe in terms of massless excitations with Planckian
effects. First of all, we show that at a classical level, the cosmological constant $\Lambda>0$ can be obtained only for
$T\rightarrow 0$. Similarly to the black hole case, when quantum effects are taken 
into account, a representation for $\Lambda$ is possible in terms of massless excitations,
provided that quantum corrections to the Misner-Sharp mass are considered. Moreover, thanks to quantum fluctuations, an effective cosmological constant arises depending on the physical scale under consideration, thus representing a 
possible solution to the cosmological constant problem without introducing a 
quintessence field. The smalness of the actual value for $\Lambda$ can be due to the existence of a quantum decoherence scale above the 
Planck length such that
the spacetime evolves as a pure de Sitter universe with a small averaged cosmological constant frozen in the lowest energy state.
 
\end{abstract}
PACS Number(s): 95.36.+x, 98.80.-k, 04.60.Bc, 05.20.-y, 04.60.-m

\section{Introduction}

The nature of the dark energy, representing about the $68\%$ of the actual energy content of the universe, is perhaps one of the biggest issue in modern physics.
In general relativity, dark energy is depicted in terms of the cosmological constant $\Lambda$, with energy density
${\rho}_{\Lambda}=\frac{\Lambda c^2}{8\pi G}$ and constant equation of state $p_{\Lambda}=-c^2{\rho}_{\Lambda}$. The physical origin 
of $\Lambda$ is still obscure. Since the cosmological constant enters in Einstein's equations as 
$T_{\mu\nu}=-\Lambda g_{\mu\nu}$,
its natural explanation is in term of a vacuum energy. Nevertheless, naive quantum field theory 
computations suggest for $\Lambda$ a value looking like $\Lambda\sim\frac{1}{L_P^2}$, with $L_P$ the Planck length, a value about
$10^{122}$ orders greater than the one really observed
\cite{5,6,7,8,9,10}: this is named 
'vacuum catastrophe'. In fact, after introducing a cutoff energy scale $E_c$, we have ${\rho}_{vac}\sim E_c^4$ with the 
effective measured
cosmological constant $\overline{\Lambda}$ and the bare one $\Lambda$ related by
\begin{equation}
\overline{\Lambda}=\Lambda+{\Lambda}_{vac}, 
\label{1}
\end{equation}
where ${\Lambda}_{vac}=8\pi G{\rho}_{vac}$. In order to have the observed value $\overline{\Lambda}=10^{-52}/m^2$, a magic cancellation
of about $10^{122}$ orders is required, but no physically realistic realizations have been yet obtained and serious doubts exist on the possibility that such a cancellation is physically possible according to general relativity. Moreover, supersymmetry is required,
but this beautifull mechanism has not been at present day observed at LHC.\\
In the literature, to overcome these issues, a lot of dark energy models with a 
time-dependent equation of state $p(t)=-\gamma(t)\rho(t) c^2$ with $\gamma>1/3$ have been proposed based principally on a modification of the
general relativity at infrared scales ($f(R)$ models \cite{12,13,14}) or adding ad hoc potentials to the energy-momentum tensor.
However, at present day, the description of the dark energy in terms of the cosmological constant (the simplest field satisfying
the Klein-Gordon equation) within the general relativity is still lacking. In any case, the physical origin of
$\Lambda$, its smallness and the role of the vacuum energy in the dynamic of the universe are yet fundamental unsolved problems.

Recentely, I have proposed \cite{1,2,3,4} a way to depict the black hole entropy in terms of trapped gravitons \cite{1,2} together with 
the logarithmic entropy corrections \cite{3}. The treatment has been extended in \cite{4} to any massless excitation. 
In particular, in \cite{3,4}  
a possible mechanism transforming a radiation field into a one with a linear equation of state is presented as a theorem. In \cite{3}
it has been shown that this mechanism naturally gives, after considering quantum gravity corrections, the well known logarithmic corrections to the black hole entropy. In \cite{4}, it has been shown that, contrary to the black hole case, macroscopic configurations with an equation of state with $\gamma=-1$ are possible in the static case. In this paper,  
we will generalize the treatment in \cite{4} to a cosmological context. Our approach is phenomenological in the sense that 
no underlying quantum gravity proposal is used, but our reasonings are based on sound arguments concerning
general relativity and quantum mechanics. \\
In section 2 we present our setups together with a generalization of the theorem present in \cite{4}. In section 3 we depict the bare cosmological constant in terms of massless excitations. In section 4 we dress the bare cosmological constant by considering quantum corrections, while section 5 is devoted to a study of the origin of the cosmological context together with the crossover to 
classicality.
Finally, section 6 collects some conclusions and final remarks.

\section{The model: massless excitations with a linear equation of state in de Sitter universe}

As a  first step, we must have at our disposal a suitable expression for the (quasi-local) energy
in  Friedmann spacetimes in comoving coordinates:
\begin{equation}
ds^2=-c^2dt^2+\frac{a^2(t) dr^2}{1-kr^2}+a^2(t)r^2\left(d\theta^2+\sin^2\theta\;d\phi^2\right).
\label{2}	
\end{equation}
In the background (\ref{2}), as well known, this quasi-local energy, 
reducing in the Newtonian limit to the matter-energy content of the spacetime together with its
gravitational energy, 
is provided by the Misner-Sharp mass $M_{ms}$ \cite{15} with associated energy
$E_{ms}=M_{ms}c^2$ . After defining $L=a(t)r$ we have:
\begin{equation}
E_{ms}(L)=\frac{c^4 L}{2G}\left(1-h^{ab}{\partial}_aL\;{\partial}_b L\right),
\label{3}
\end{equation}
where $h^{ab}$ is the two metric spanned by the coordinates $(t,r)$. In terms of the apparent horizon $L_A$ with 
$L_A=\frac{c}{\sqrt{H^2+\frac{k}{a^2(t)}}}$, where $H$ denotes the Hubble rate, we have:
\begin{equation}
E_{ms}=\frac{c^4}{2G}\frac{L^3}{L_A^2}.
\label{4}
\end{equation}
Expression (\ref{4}) can be seen as the quai-local energy within a proper volume $V(L)=4\pi L^3/3$.
We are interested in a de Sitter flat expanding universe obtained from (\ref{2}) with $k=0$ and
$a(t)=e^{ct\sqrt{\frac{\Lambda}{3}}}$ together with 
$H(t)=H_{\Lambda}=c\sqrt{\frac{\Lambda}{3}}$.\\
After specifying the expression for the quasi-local energy within a volume of proper areal radius $L$,  we must specify the macroscopic model.
To this purpose, consider a spherical region of proper volume $V(L)$: the contributions to $E_{ms}$ in (\ref{4}) are 
supposed to be provided by 
massless modes (perhaps gravitons or also photons)
with proper wavelengths $\lambda$ such that $\lambda\leq L$. As a consequence, 
concerning the allowed discrete spectrum, for the angular
frequencies ${\omega}^{(0)}$ we can use the following expression, in line with \cite{4}:
\begin{equation}
{\omega}^{(0)}_n=a\frac{cn}{L},\;a\in{\Re}^+,\;\;n\in{\bf N}.
\label{5}
\end{equation}
Note that, when quantum fluctuations are taken into account in section 4, we have a natural ultraviolet cutoff given by the 
Planck length $L_P$, with $L\geq L_P$.\\
We can identify $E_{ms}$ as the internal energy $U^{(0)}$
of massless excitations inside \footnote{In \cite{16} to a Friedmann flat spacetime at
the apparent horizon we associated a zero internal energy. This result is not in disagreement with the asumption of this paper. In fact, in \cite{16}, the internal energy appearing in the first law at the apparent horizon is 
a summation bewteen the Misner-Sharp mass and the negative contribution due to the non static dinamic of the universe and these contributions cancel out.}. With the (\ref{5}), we can calculate the partition function
$Z_T^{(0)}={Z^{(0)}}^N$ for $N$ excitations:
\begin{equation}
Z^{(0)}=\sum_{n=0}^{+\infty} e^{-\beta\hbar{\omega}^{(0)}_n}=
\frac{1}{1-e^{-\frac{ac\hbar\beta}{L}}},\;\;\beta=\frac{1}{K_BT},
\label{6}
\end{equation}
where $T$ is the temperature and $K_B$ the Boltzmann constant. 
As usual we have $U^{(0)}=-{\left(\ln Z_T\right)}_{,\beta}$ (comma denote partial derivative):
\begin{equation}
U^{(0)}=\frac{ca\hbar N}{L\left(e^{\frac{ac\hbar\beta}{L}}-1\right)},
\label{7}
\end{equation}
with the constraint
\begin{equation}
U^{(0)}=\frac{c^4}{2G}\frac{L^3}{L_A^2}=\frac{ca\hbar N}{L\left(e^{\frac{ac\hbar\beta}{L}}-1\right)}.
\label{8}
\end{equation}
Note that in the classical limit $L>>ac\hbar\beta$, we obtain the classical result 
$U^{(0)}=NK_B T$.
For the free energy we have $F_T=-NK_BT\ln(Z^{(0)})$ with the pressure $p^{(0)}$ given by
${F_T}^{(0)}_{,V}=-p^{(0)}$. From (\ref{6}) and (\ref{7}) we obtain
$p^{(0)}V=\frac{U^{(0)}}{3}$, i.e. as expected for (\ref{5}) a radiation field \cite{3,4}.  
In order to depict the cosmological constant equation of state $p_{\Lambda}=-c^2{\rho}_{\Lambda}$, we must obtain the suitable frequency
spectrum. As in \cite{3,4}, we can write, without loss of generality, the frequency $\omega$ in the form
\begin{equation}
\omega(n,L)={\omega}^{(0)}_n+\frac{\Phi(L)}{N},
\label{9}
\end{equation}
where $\phi(L)$ is a function to be determinated. Moreover, we have 
$F_T=-N K_B T\ln(Z_T)=F_T^{(0)}+\hbar\;\Phi(L)$.
It is worth to be noticed that, in practice, the added term due to $\Phi$ it gives a contribution to the energy that is independent on the 
temperature $T$. This is reminescent of solid state physics, where for the free energy, to the summation over the oscillation frequencies
of the atoms, a further term is added, namely ${\epsilon}_0(N/V)$ independent on the temperature and depending only on the 
density of the solid: this added term depicts the energy of the atoms of the solid in the equilibrium configuration.
In order to describe the cosmological constant in a physically sound way within general relativity, the energy density must be constant in time and space and given by ${\rho}_{\Lambda}=\frac{\Lambda}{8\pi G}$. In fact, this is what we obtain by taking 
${\rho}_{\Lambda}=E_{ms}/V$, with $E_{ms}$ given by (\ref{4}). The next step is to adopt 
and modify the theorems quoted in \cite{3,4} in
a form suitable for a cosmological context. To this purpose, since the energy is fixed by the general relativistic expression 
(\ref{4}) and as a result we have at our disposal the expression for the internal quasi-local energy $U(L)$, the point is to obtain the exact expression for the spectrum of the excitations given by (\ref{9}) leading to $U(L)$ via the partition function 
$Z_T={\left(\sum_{n=0}^{\infty}e^{-\beta\hbar\omega(n,L)}\right)}^N$. The following result still holds:

{\bf Proposition:} {\it Let ${\omega}_0$, given by the (\ref{5}), denote the angular frequency of $N$ 
masless excitations within a volume of proper areal radius $L$. The excitations with energy
$\hbar\omega=\hbar{\omega}_0+\hbar\frac{\Phi(L)}{N}$ have a linear equation of state
$PV=\gamma U$ provided that the differentiable function $\Phi(L)$
satisfies the following equation} 
\begin{equation}
\hbar\left[L\;{\Phi}_{,L}(L)+\Phi(L)\right]=U(L)(1-3\gamma), \label{10}
\end{equation}
{\it together with the condition}
\begin{equation}
U(L)-\hbar\;\Phi(L) > 0.
\label{11}
\end{equation}
\begin{proof}
With the usual relation $U(L)=-{\ln(Z_T)}_{,\beta}$, we obtain
\begin{equation}
U=U^{(0)}+\hbar\;\Phi(L).
\label{13}
\end{equation}	
Since from the (\ref{7}) we have $U^{(0)}>0$, condition (\ref{11}) follows. For the free energy we have
$F_T=-N K_B T\ln(Z_T)=F_T^{(0)}+\hbar\;\Phi(L)$. Moreover
\begin{equation}
{F_T}_{,V}=\hbar\;{\Phi}_{,L}\;L_{,V}+L_{,V}\;{F_T^{(0)}}_{,L}=-P,	
\label{14}
\end{equation}	
with $L_{,V}\;{F_T^{(0)}}_{,L}=-P^{(0)}$ and $P^{(0)}V=\frac{U^{(0)}}{3}$. Hence, from (\ref{14}) we get
\begin{equation}
\hbar\frac{L}{3V}\;{\Phi}_{,L}-\frac{U^{(0)}}{3V}=-P.	
\label{15}
\end{equation}	
After using the (\ref{13}) with $PV=\gamma U(L)$, from (\ref{15}) we obtain the equation (\ref{10}). 
\end{proof}
Note that for a radiation fiels, i.e. $\gamma=1/3$, the right hand side of (\ref{10}) becomes zero. The solution
$\Phi(L)\sim 1/L$ is solution of the homogeneous equation (\ref{10}) giving a radiation field equation of state. 
By requiring that for $\gamma=1/3$ the spectrum is provided by the (\ref{5}), we can set to zero the homogeneous solution of the
(\ref{10}). An important point of the present approach is provided by equation (\ref{8}). In fact, the spectrum 
(\ref{5}) is a microscopic description of the matter-energy content of the spacetime that in turn is fixed, 
at a semi-classical level\footnote{Without quantum gravity motivated corrections considered in section 4},
by the Misner-Sharp energy (\ref{3}); this fact is often missing in the literature.\\
It is interesting that the so obtained proposition is in fact independent on the explicit form of the potential
$U^{(0)}$ and also on the explicit (discrete) expression for ${\omega}^{(0)}_n$ in (\ref{5}). The fundamental assumption is that 
${U}^{(0)}$ represents a radiation field. Thus, in principle, also massless fermionic excitations representing a 
radiation field could contribute to ${U}^{(0)}$, with ${U}^{(0)}$ representing the whole contribution due to fermions and bosons and
with obviously a different expression for (\ref{6}) but with the solution 
for $\Phi(L)$ left unchanged, provided that the expression for $U(L)$ is fixed by Misner-Sharp expression.
\footnote{To this purpose, in principle also massive fermions can be considered.}.\\
However, we stress that our approach for the cosmological constant is different from usual ones present in the literature and based 
principally on supersymmetry. As will be shown in section 4, the equation of state for the cosmological constant only emerges when
Planckian fluctuations come into action, without introducing the usual vacuum made of all kind of particles.   
Hence, 
we refer to "vacuum" as Planckian fluctuations, and massless excitations seem to be more appropriate at Planckian scales.\\
With (\ref{10}), we can now explore suitable solutions for the cosmological constant case.

\section{Bare cosmological constant with massless thermodynamically dead excitations}

To start with, we must solve the equation (\ref{10}) with $U(L)=E_{ms}(L)=\frac{c^4}{2G}\frac{L^3}{L_A^2}$. We obtain:
\begin{equation}
\hbar\Phi(L)=(1-3\gamma)\frac{c^4}{8G}\frac{L^3}{L_A^2}.
\label{16}
\end{equation}
With the solution (\ref{16}), the existence condition (\ref{11}) it gives $\gamma >-1$ \cite{3,4}. This fact clearly shows that
the case $\gamma=-1$ suitable for a cosmological constant is a very special one. The only possibility to obtain the cosmological constant case in our semi-classical background (no quantum gravity effects) is to set $T=0$ for our system of massless excitations. Hence, the cosmological de Sitter spacetime with a positive cosmological constant $\Lambda$ can be depicted in terms of massless excitations 
at the absolute zero temperature. This fact is in agreement with the idea that the cosmological constant is a manifestation of vacuum energy.
In fact, for $T=0$ we have $U^{(0)}(L)=0$ and consequently $U(L)=E_{ms}(L)=\hbar\Phi(L)$, with $\Phi(L)$ given by
(\ref{16}) with $\gamma=-1$ and independent on the temperature $T$. For $T=0$, according with the third law of thermodynamics, we have
zero entropy $S(T=0)=0$.
It is worth to be noticed that $T\rightarrow 0$ represents the lowest energy state for the system: in such a state we have a non-vanishing
energy for the spacetime. This fact is in complete agreement with Heisenberg uncertainty relation dictating that a quantum system at
the lowest energy state cannot have a vanishing energy: this non-vanishing energy within 
a given proper volume $V$ is the Misner-Sharp one $E_{ms}$ representing
a non vanishing cosmological constant, exactly what physically we expect. For the reasonings above, in the following we denote with
$\Lambda=\Lambda(T=0)$ the 'bare' cosmological constant. The calculations of this section clearly show that, in order to understand
the true nature of the cosmological constant, quantum Planckian corrections must be considered.
In the next section we will 'dress' $\Lambda$ by introducing quantum gravity
motivated modifications to the expression (\ref{8}) for $U(L)$.

\section{Cosmological constant with quantum gravity motivated corrections}

In this section we depict the cosmological constant by considering corrections due to quantum fluctuations that are expected in 
a quantum gravity regime. Our approach is in some sense 'phenomenological'. This means that, in order to apply the
(\ref{10}), we must obtain an expression for $U(L)$ with quantum-gravity motivated corrections. We may suppose that at a 
quantum gravity level the spacetime cannot longer be depicted with a classical metric as the (\ref{2}), but the
spacetime granularity comes into action. There, a quantum spacetime \cite{17} motivated by the non-commutativity of the spacetime
coordinates can be quoted. Physically motivated spacetime uncertainty relations (STUR) can be found in \cite{16} in a Newtonian approximation 
and further generalized in \cite{18} and in \cite{19} in a Friedmann flat background. In particular, in \cite{19}
physically motivated STUR have been obtained in spatial Cartesian coordinates $\{x^i\}$  by introducing proper coordinates given by
${\eta}^i= a(t) x^i$. In the coordinates ${\eta}^i$, the Heisenberg uncertainty relations can be written, in any allowed state 
$s$, as
\begin{gather}
\Delta_s E\Delta_s t\geq\frac{\hbar}{2},\qquad
\Delta_s p_{{\eta}^i}\Delta_s{\eta}^i\geq\frac{\hbar}{2},\;\;\forall i=1,2,3.
\label{17}
\end{gather}
STUR in a Friedmann flat quantum spacetime \cite{19} can be obtained in terms of the ratio  $\Delta_s A/\Delta_s V$, where,
$\Delta_s A=\sum_{i,j, i\leq j}\Delta_s {\eta}^i \Delta_s {\eta}^j$ and 
$\Delta_s V=\prod_i\Delta_s {\eta}^i$:
\begin{eqnarray}
& & \frac{\sqrt{\Delta_s A}}{4\sqrt{3}}+\frac{s(H)\Delta_s A}{12c}\geq \frac{L^2_P}{2} 
\frac{\Delta_s A}{\Delta_s V}, \label{18}\\
& & c\Delta_s t\left(\frac{\sqrt{\Delta_s A}}{4\sqrt{3}}+\frac{s(H)\Delta_s A}{12c}\right)\geq \frac{L^2_P}{2}.\label{19}
\end{eqnarray}
For the spacetime (\ref{2}) with $k=0$
representing a de Sitter expanding universe, the Hubble rate $H$ is constant, and as a consequence the mean value $s(H)$ 
is a constant in a de
Sitter spacetime. Moreover, thanks to quantum effects, in (\ref{18}) and (\ref{19}) we can substitute the bare cosmological constant
$\Lambda$ with the one 'dressed' by quantum interactions $\overline{\Lambda}$, i.e. 
$s(H_{\overline{\Lambda}})=c\sqrt\frac{\overline{\Lambda}}{3}$. Note that, since  $1/\sqrt{\overline{\Lambda}}$ is of the order of the
dimension of the apparent horizon (Hubble radius of the universe)
the following condition is expected to hold:
\begin{equation}
s(H_{\overline{\Lambda}})\sqrt{\Delta_s A} \leq c.
\label{20}
\end{equation}
Moreover, the state such that the STUR (\ref{18})-(\ref{19}) are satured are called maximal localizing states \cite{17}: they are 
state (spherical) with all uncertainties of the same magnitudo, i.e. 
$c\Delta_s t\sim \Delta_s {\eta}^i\sim \Delta_s {\eta}$. Hence, the STUR (\ref{18})-(\ref{19}) do imply that 
$\Delta_s {\eta}^i\geq \chi L_P$, with $\chi$ of the order of unity. As a consequence, the (\ref{17}) for $\Delta_s E$ becomes
$\Delta_s E\geq\chi\frac{c\hbar}{2\Delta_s {\eta}}$.
In terms of our proper variables $L=a(t)r$ we have $\Delta_s L\sim \Delta_s {\eta}^i$ 
and thus \footnote{Note that in the spherical case the STUR implies that $\Delta_s L\geq L_P$, thus representing an ultraviolet cutoff.}
\begin{equation}
\Delta_s E\geq\chi\frac{c\hbar}{2\Delta_s L}.
\label{21}
\end{equation}
In this section we treat the parameter $\chi$ as a constant that it is expected of the order of unity. Expression (\ref{21}) physically motivates the following expression for 
$U(L)$ in a proper volume $V=4\pi L^3/3$ dressed by quantum fluctuations:
\begin{equation}
U(L)=\frac{c^4}{2G}\frac{L^3}{L_A^2}+\chi\frac{c^4}{2G}\frac{L_P^2}{L}.
\label{22}
\end{equation}
The next step is to use expression (\ref{22}) for $U(L)$ in (\ref{10}) with the solution
\begin{equation}
\hbar\Phi(L)=(1-3\gamma)\frac{c^4}{8G}\frac{L^3}{L_A^2}+\frac{\chi(1-3\gamma)c^4}{2G}
\frac{L_P^2}{L}\ln\left(\frac{L}{L_0}\right),
\label{23}
\end{equation}
with $L_0$ a positive constant. For $\chi>0$, the condition (\ref{11}) becomes:
\begin{equation}
1-(1-3\gamma)\ln\left(\frac{L}{L_0}\right)>0.
\label{24}
\end{equation}
We are interested in the cosmological constant case  $\gamma=-1$ with the solution $L<L_0 e^{\frac{1}{4}}$. Since in our model we have 
a natural constant $\overline{\Lambda}$, we may take $L_0\sim 1/\sqrt{\overline{\Lambda}}\sim L_A$ and as a consequence $L$ can be taken also of the order of Hubble radius. As we will show in the next section, in order to be in agreement with general relativity
\footnote{In general relativity, in a Friedmann flat cosmology (\ref{2}), a fluid with an equation of state with $\gamma=-1$
must have a constant energy density, namely ${\rho}_{\overline{\Lambda}}$.}
without introducing quintessence exotic dark energies, the proper radius $L$ must be interpreted as an effective physical length-scale.\\
Another possibility, explored in \cite{4}, is to take $\chi<0$. In this case a macroscopic configuration is possible with a minimum radius
$L_0$ with $L>0$, but with $L_0=sL_P$ and $s$ of the order of unity or greater. However, as we will show in the next section, 
for our interpretation of the dependence on $L$  of $\overline{\Lambda}$, where the range of $L$ is requested at the Planck length up 
to macroscopic scales where decoherence and classicality are expected to hold, the proposal $\chi>0$ 
is certainly the most appropriate.\\
Regarding the question of the temperature, in the section above we have associated to a classical spacetime with non-vanishing bare cosmological constant $\Lambda$ a vanishing temperature $T=0$. This result is in agreement 
with physical reasonability for a bare cosmological constant. However, it is customary to associate to the apparent horizon of a given
classical  Friedmann universe a temperature $T_h$, namely $T_h=\frac{c\hbar}{4\pi K_B L_A}$. This temperature looks like an Unruh temperature for an 
ingoing radiation from the apparent horizon. In \cite{16} it has been shown that, in a de Sitter universe, since the apparent horizon is static, one can associate to the dark energy a temperature given by $T_h$. The point is: it is this temperature the one effectively measured by a thermometer or is a temperature arising to satisfy the first law of thermodynamics ? It is not easy to ask to this question in a
rigorous way. For example, in \cite{20} it has been advanced the possibility that the Unruh temperature is not the one measured by a thermometer,
thus it does not represent an exchange of heat with a surrounding gas, but rather it is caused by quantum effects generated by a local coupling 
between the thermometer and the vacuum state. When cosmological constant is dressed with quantum corrections, one expect that a non-zero temperature $T_{\overline{\Lambda}}$
can arise. For the reasonings above and present in \cite{20}, this temperature $T_{\overline{\Lambda}}$ could be different from
$T_h$. If we are wilings to accept, thanks to the holographic principle, that to a dark energy is associated a non-vanishing entropy 
proportional to the area of the apparent horizon\footnote{See for example \cite{1,2,16,21} and references therein}, then we can set
$T_{\overline{\Lambda}}=b T_h$, with $b\in\Re^+$ of the order of unity or less. These reasonings are in agreement, on general
grounds, with the first law of thermodynamic in the usual form 
\begin{equation}
T_{\overline{\Lambda}}dS_{\overline{\Lambda}}=dU+PdV.
\label{25}
\end{equation} 
In fact, the cosmological constant equation of state requires that, by inspection of (\ref{25}),
$T_{\overline{\Lambda}}dS_{\overline{\Lambda}}=0$: for the bare case with  $T_{\Lambda}=S_{\Lambda}=0$ this equation is trivially
satisfied. For  $T_{\overline{\Lambda}}>0$, we must have $S_{\overline{\Lambda}}=k$, with $k\in\Re^+$. Since apparent horizon of 
a de Sitter cosmological universe is constant in time, the entropy of a de Sitter spacetime is expected to be constant in time
at the apparent horizon.
With and only with the choice $T_{\overline{\Lambda}}=b T_h$ we have an entropy proportional to the area of the apparent horizon together 
with logarithmic corrections, thanks to the logarithmic term in (\ref{21}). In this way, the entropy of a de Sitter spacetime 
at the apparent horizon becomes the usual one for a black hole with the expected Planckian-fluctuations
motivated logarithmic corrections.\\ However, the explicit form for $T_{\overline{\Lambda}}$ with the related expression for the entropy play 
a marginal role in the paper.  

\section{An effective length-scale depending cosmological constant}

In this section we study the formula (\ref{22}) for the quasi-local energy inside the region of proper areal radius $L$. Formula 
(\ref{22}) implies that
\begin{equation}
{\rho}_{\overline{\Lambda}}=\frac{c^2\Lambda}{8\pi G}+\frac{3\chi c^2}{8\pi G}\frac{L_P^2}{L^4},
\label{26}
\end{equation}
that in turn implies
\begin{equation}
\overline{\Lambda}=\Lambda+\frac{3\chi L_P^2}{L^4}.
\label{27}
\end{equation}
It should be stressed that the meaning of (\ref{27}) is that the observable cosmological constant $\overline{\Lambda}$
is composed of two contributions: the former due to $\Lambda$ that is the contribution of the radiation massless field
\footnote{In sections above we have named this contribution bare cosmological constant since it can be also seen
as the cosmological constant at $T=0.$}, while the latter is 
the one dressed by Planckian effects. As stated by formula (\ref{10}), this term is fundamental in order to have the equation of
state suitable for a cosmological constant.\\
A naive interpretation of the formula (\ref{27}), since of the dependence on $L$, is in terms of quintessence dark energy. A
quintessence model requires a variating equation of state, while in our setups the equation of state is constant in time and space
with $\gamma=-1$. To obtain a physical understanding of the (\ref{19}), it should be noted that, thanks to the results of sections
3 and 4, the cosmological constant case with $\gamma=-1$ can be obtained only thanks to quantum Planckian fluctuations in the
'phenomenological' formula (\ref{22}) of the order of $\sim 1/L$. Hence, the proper-length parameter $L$ can be correctly interpreted as 
a parameter depicting the scale at which the physics is considered (averaged). In fact, we may consider the spacetime metric
$h_{ab}$ as a fluctuating quantity depending on general spatial 
coordinates $\{x^a\}$ and a time coordinate $\overline{t}$, with line element
\begin{equation}
ds^2=-c^2 d{\overline{t}}^2+h_{ab}(\overline{t}, x^a)dx^a dx^b.
\label{28}
\end{equation}
First of all, the time $\overline{t}$ could be considered as an average time of fluctuating metrics on a given spherical box of proper areal radius $L$. We can assume that when classicality is recovered,
the time $\overline{t}$ is nothing else but the cosmic one $t$
present in the classical metric (\ref{2}): in a quantum spacetime $t$ becomes \cite{19} an essentially self-adjoint
operator\footnote{As shown in \cite{19}, in a spacetime with big bang, $t$ can be symmetric with a unique self-adjoint
extension, while in a de Sitter spacetime with no big bang $t$ can be self-adjoint.} 
satisfying the STUR (\ref{18}) and (\ref{19}). Concerning the spatial metric $h_{ab}$, we can adopt, after some suitable
adjustment, the well known Buchert formalism \cite{22}. There, we can consider a slice of constant $\overline{t}$
with $\overline{t}=s(t)=k\in\Re$, with $s$ a quantum allowed state with respect to the (\ref{18})-(\ref{19}).\\
To start with, consider a proper volume $V(L)$.
For any scalar quantity $\psi(s(t), x^i)$, the average with respect to the volume $V(L)$ is:
\begin{equation}
{<\psi(s(t),x^i)>}_{V(L)}=\frac{1}{V_{L}}\int_{V(L)}\psi(s(t),x^i)\sqrt{g^{(3)}}d^3x,
\label{29}
\end{equation}
where $g^{(3)}$ denotes the determinant of the three metric $h_{ab}$ on the slice a $s(t)=const.$ and
\begin{equation}
V_{L}=\int_{V(L)}\sqrt{g^{(3)}}d^3x.
\label{30}
\end{equation}
Moreover, for the averaged expansion rate ${<\theta>}_{V(L)}$ we have
\begin{equation}
{<\theta>}_{V(L)}=\frac{{\dot{V}}_{L}}{V_{L}}=
3\frac{{\dot{a}}_{V(L)}}{a_{V(L)}},
\label{31}
\end{equation} 
where dot denotes time derivative with respect to $\overline{t}$ and
the dimensionless effective scale factor $a_{V(L)}(s(t))$ is given by
\begin{equation}
a_{V(L)}(s(t))={\left(\frac{V_{L}(s(t))}{V_{L}(s(t_0))}\right)}^{\frac{1}{3}},
\label{32}
\end{equation}
with $s(t_0)$ an initial time. For the averaged Hubble flow we have
$H=\frac{{<\theta>}_{V(L)}}{3}$. Since on average, in order to obtain the de Sitter 
spacetime, we expect a spatiallty flat metric, we can set to zero the averaged curvature
$\mathcal{R}$. As a consequence, we have an effective averaged metric given by:
\begin{equation}
ds^2=-c^2 d{\overline{t}}^2
+a_{L}^2(s(t))\left[dr^2+r^2\left(d\theta^2+\sin^2\theta d\phi^2\right)\right],
\label{33}	
\end{equation}
where we used the more short notation $a_{V(L)}(s(t))=a_L(s(t))$.
With the above definition, the relevant equations for our purposes
for an irrotational cosmological constant fluid at the fixed averaged scale $L$ for the above modified 
Buchert equations are:
\begin{eqnarray}
& &3\frac{{\dot{a_L}}^2}{a_L^2}=
c^2\Lambda+\frac{3c^2\chi L_P^2}{L^4}
-\frac{\mathcal{Q}_L}{2},\label{34}\\
& &\mathcal{Q}_L=\frac{2}{3}\left[<{\theta}^2>-{<\theta>}^2\right]-2{<\sigma>}^2.\label{35}
\end{eqnarray}
In (\ref{35})  $\mathcal{Q}_L$ is the kinematical backreaction, while $\sigma$ represents the shear. 
In a cosmological context \cite{23} the shear is negligible on scales also smaller than the scale of homogeneity.
In a similar manner, in our context where the scales are Planckian or microscopic, 
it is reasonable that on scales soon after the Planck one this term is also negligible. Moreover, the kinematical
backreaction term  $\mathcal{Q}_L$ is also expected to give small \cite{23} values and as a consequence it gives 
a small contribution for the cosmological constant. Under these assumptions, formal integration of (\ref{34}) at the scale
$L$ it gives
\begin{equation}
a_L(s(t)=a_L(s(t_0))e^{\left(c\int_{s(t_0)}^{s(t)}\sqrt{\frac{\Lambda_{eff}}{3}}d\overline{t}\right)},
\label{36}
\end{equation}
with
\begin{equation}
\Lambda_{eff}=\Lambda+\frac{3\chi L_P^2}{L^4}-\frac{\mathcal{Q}_L}{2}.
\label{37}
\end{equation}
The term $\frac{\mathcal{Q}_L}{2}$ in (\ref{37}) 
as stated above, is expected to be negligible on 
length scales above the Planck one, and also is expected to rapidely decreases for time-scales $t>t_p$, with $t_p$
the Planck time. In fact, the integrability condition for the system (\ref{34})-(\ref{35}) is
\begin{equation}
6\mathcal{Q}_L\dot{a_L}+a_L\dot{\mathcal{Q}_L}=0,
\label{38}
\end{equation}
showing that $\mathcal{Q}_L$ scales as $\mathcal{Q}_L\sim a_L(s(t_0))/a_L(s(t))^6$.
The (\ref{36}) depicts a new view to look to the cosmological constant problem: instead of a 
naive summation over the  different vacuum energy contributions (Planck scale, QCD scale...), the cosmological constant is provided from
an average over a given physical length scale $L$. For a bigger and bigger $L$, the vacuum contribution proportional to
$1/L^4$ becomes smaller and smaller and for $L>>L_P$ it becomes practically negligible thus representing a solution, at least, to the 
'old' cosmological constant problem.\\
It should be noticed that, in the usual cosmological background, there exists a scale $L_o$, named scale of homogeneity, such that  
an average on scales $L$ greater than $L>L_o$ is obviously trivial since the metric factor $a_L$ is in fact no longer dependent on 
$L$. What means this reasonings translated in the language of an effective quantum gravity theory
\footnote{It must be stressed again that our approach is phenomenological in the sense that no underlying quantum gravity theories
are advanced and our procedure is not a proposal to quantize the gravity.}? On general grounds, for a quantum system, it is expected that 
a decoherence scale emerges (see for example \cite{24,25} in a quantum gravity context) and so also this scale is expected to arise 
for the quantum gravity regime. This is obviously a complicated matter and to treat the problem 
in a suitable manner we need the quantum gravity theory that 
it is not actually at our disposal. In particular we may think to a decoherence length $L_D$ such that for $L>L_D$
quantum Planckian corrections become irrelevant and the 
transition to classicality comes into action. At such a scale, the term $\mathcal{Q}_{L_D}$ in (\ref{36}) can be neglected an as a result a de Sitter phase emerges with
\begin{equation}
a_{L_D}(s(t)\sim a_{L_D}(s(t_0))e^{c(s(t)-s(t_0))\sqrt{\frac{\overline{\Lambda}}{3}}},
\label{39}
\end{equation}
where at the transition at the classicality we have $s(t)=\overline{t}=t$ and $\overline{\Lambda}$ given by 
\begin{equation}
\overline{\Lambda}=\Lambda+\frac{3\chi L_P^2}{L_D^4}.
\label{40}
\end{equation}
After performing an average on scales $L>L_D$, since the classicality is reached, the average acts trivially as obviously acts on the 
classical metric (\ref{2}).\\
In the new approach presented in 
this paper to the cosmological constant problem, the smallness of $\overline{\Lambda}$ is due to the existence of a decoherence scale
$L_D$. In our phenomenological
approach, some numerical examples can de done.\\
To start with, it is interesting to calculate the scale $L_D$ in such a way that the term
$\frac{3\chi L_P^2}{L_D^4}$ is of the 
same magnitudo expected for the cosmological constant, i.e. $\sim 10^{-52}/m^2$: we found, after taking 
$\chi$ of the order of unity, $L_D\sim 10^{-5} m$. This could look as a rather big value, but it should be noted that decoherence in a gravitational field can be rather huge \cite{24}. In any case, in our calculations we have supposed that the constant $\chi$ present
in (\ref{22}) is the same as the one present in the STUR (\ref{21}). Hence, we can alleviate this assumption and suppose that the effective 
energy $U$ is provided by
\begin{equation}
U(L)=\frac{c^4}{2G}\frac{L^3}{L_A^2}+\xi\frac{c^4}{2G}\frac{L_P^2}{L},
\label{41}
\end{equation}
where $\xi\in(0,1)$.
Note that, since the measured cosmological constant is not $\Lambda$ but $\overline{\Lambda}$, the term $L_A^2$ in the semiclassical term
of $U$ in (\ref{41}) can be confusing. To avoid any possible confusion for the reader, the following changement can be consistently given.
First of all, we introduce a new parameter, namely $\Gamma$ and we substitute $L_A^2$ in (\ref{41}) with
$L_{A}^2=\frac{3}{\Gamma}$: in practice we have made the substitution $\Lambda\rightarrow\Gamma$ in order to stress that $L_A$ refers to a bare 
cosmological constant $\Gamma$. 
The first term in (\ref{41})
is the one due to a massless radiation field with constant density
${\rho}_r=\frac{c^2\Gamma}{8\pi G}$ and the second one is due to Planckian fluctuations. The energy expression (\ref{41}) depicts the physics of
our system at and above Planck scale with the cutoff $L_{inf}\sim L_P$ dictated by the STUR (\ref{18})-(\ref{19}). As far as the classicality is 
reached, the averaged metric evolves according equation (\ref{39}). 
For $\overline{\Lambda}$ from (\ref{41}) we obtain
\begin{equation}
\overline{\Lambda}=\Gamma+\frac{3\xi L_P^2}{L^4}.
\label{45}
\end{equation}
How can we physically mark the crossover to the classicality ?
Quite remarkably, our phenomenological model can provide a physically sound answer. To this purpose, we expect that at the
critical decoherence scale-length $L_D$ the quantum system undergoes a transition phase and the value of the measured dressed
cosmological constant $\overline{\Lambda}$ remains, for length-scales $L>L_D$ frozen to the value at $L_D$, 
i.e. for $L>L_D\rightarrow \overline{\Lambda}(L)=\overline{\Lambda}(L=L_D)$. The point $L=L_D$ must 
thus be an absolute minimum for 
$U(L)$, i.e. the configuration at the length-scale $L=L_D$ is at the lowest state enegy configuration:
\begin{equation}
L_D={\left(\frac{\xi L_P^2}{\Gamma}\right)}^{\frac{1}{4}}.
\label{46}
\end{equation}
Thanks to (\ref{46}), for $\overline{\Lambda}(L=L_D)$ we get:
\begin{equation}
\overline{\Lambda}(L=L_D)=\Gamma+\frac{3\xi L_P^2}{L_D^4}=4\Gamma.
\label{47}
\end{equation}
Equation (\ref{47}) relates the phenomenological parameter $\Gamma$ to the measured cosmological constant. As a consistence check, we expect
that for $L\geq L_D$ the system evolves with the cosmological constant $\overline{\Lambda}$ and with the classical expression of the
Misner-Sharp energy. In fact, thanks to (\ref{46}) and (\ref{47}) we have:
\begin{equation}
\frac{c^4 L_D^3}{2G L_{\overline{\Lambda}}^2}=
\frac{c^4}{2G}\frac{L_D^3}{L_{A}^2}+\xi\frac{c^4}{2G}\frac{L_P^2}{L_D},
\label{48}
\end{equation}
where
\begin{equation}
L_{\overline{\Lambda}}^2=\frac{3}{\overline{\Lambda}}=\frac{3}{4\Gamma}.
\label{49}
\end{equation}
As a cosequence of the (\ref{48}), for $L>L_D$ classicality is reached and a pure de Sitter spacetime with dressed cosmological constant
(the one we measure)
$\overline{\Lambda}$ and with the usual Misner-Sharp energy 
$E_{ms}(L)=\frac{c^4 L^3}{2G L_{\overline{\Lambda}}^2}$ arises.\\
The crossover to classicality can also be understood in a thermodynamical language. In fact, suppose that the temperature
at the length-scale $L$ is given by $T=T(L)$. For the specific heat $C=\frac{dU}{dL}$ from (\ref{41}) we obtain:
\begin{equation}
C=\frac{c^4}{2G}\left(L^2\Gamma-\frac{\xi L_P^2}{L^2}\right)\frac{dL}{dT(L)}.
\label{50}
\end{equation}
It is interesting to note that exactly at $L=L_D$, the critical length-scale, we have $C=0$. Hence, at $L=L_D$, i.e. at the 
minimum of $U(L)$, the system is thermodynamically dead and as a result in this vacuum state a pure de Sitter spacetime
emerges. Moreover, by supposing $T\sim 1/L$, as happens for a black hole or for the Unruh temperature at the apparent horizon,
we have $dT/dL<0$ and as a consequence for $L<L_D$ we have $C>0$, while for $L>L_D$ we obtain
$C<0$, according to the expectation that in a classical self gravitating system the specific heat is negative. This could represent a
new way to look to the decoherence scale in a cosmological context.\\
As a further consideration, note that expression (\ref{47}) can be written as:
\begin{equation}
\overline{\Lambda}=4\xi\frac{L_P^2}{L_D^4}.
\label{51}
\end{equation} 
The formula (\ref{51}) is interesting because it gives the measured cosmological constant ($\sim 10^{-52}/m^2$) in terms of
our parameters $L_D$ and $\xi$. The decoherence length-scale $L_D$ from quantum gravity regime to
classicality could be, at least in principle, measured in
future experiments, while $\xi$ could be calculated from models on quantum Planckian fluctuations of the energy in a 
non-commutative spacetime. In any case our phenomenological model predicts an expression for $\overline{\Lambda}$
in terms of quantum gravity motivated quantities.\\
Also note that the technology presented above can be easily extended to the case where other kind of matter-energies are present.
As an exapmle, if dust is present as in the concordance model, this contribution must be included in the modified 
Buchert equations (\ref{34})-(\ref{35}): as a result a decoherence scale $L_D$ emerges representing the crossover to
classicality and the metric of the concordance $\Lambda$CDM model is obtained.\\ 
We expect that if classicality emerges near the Planck scale, the dimensioneless parameter
$\xi$ must be $\xi<<1$.\\
A final interesting problem is the following: why a positive cosmological constant? 
Within our approach it is possible to give a reasonable answer. In fact, for a negative cosmological constant
we have a spacetime (Anti de Sitter) with a constant negative curvature with a classical quasi-local mass $E_{ql}$ within a spherical box $V(L)$
looking like $E_{ql}\sim L^3$. In order to depict this negative vacuum energy, we must have an energy $U(L)$ with a stationary point 
at the decoherence length-scale $L_D$. This can be done with $U(-\overline{\Lambda})=-U(L)$, with $U(L)$ given by (\ref{41})
together with $\overline{\Lambda}=-4\Gamma$. In this way, the energy function $U(-\overline{\Lambda})$ has a local maximum
at $L=L_D$ given by (\ref{46}), and thus $L=L_D$ does not represent a true vacuum state with the lowest energy and the system is
thus instable under quantum fluctuations and as a consequence the transition to classicality cannot happens in the manner depicted above.\\
We have presented a physically sound phenomenological model that is capable to depict the cosmological constant from a 
statistical point of view, to give a solution to the old cosmological constant problem and also to give a physical mechanism explaining
the origin of our small cosmological constant as an averaged spacetime at the length-scale $L_D$ representing the decoherence from a quantum
gravity regime to classicality. The critical length-scale $L_D$ is defined as the one giving an absolute minimum for the 
energy $U(L)$ thus representing the lowest-state energy, i.e. a true vacuum state that is frozen for scales
$L>L_D$.\\
In this regard, our approach is completely new since it indicates a more physical and serious way to treat the vacuum energy, by means of
a length-scale dependent cosmological constant in complete agreement with general relativity and quantum
mechanics.\\
An alternative interesting approach to the vacuum energy problem can be found in \cite{26}. There, the vacuum energy is 
considered fluctuating and rather inhomogeneous over the whole spacetime, whit the fluctuating spacetime obtained as a stochastic
field of inhomogeneous metrics. Parametric resonance is invoked to obtain an emerging spacetime with a small cosmological constant.
Although the physical mechanism to explain the origin of a classical de Sitter universe presented in \cite{26} is different 
from the one presented in this article, the paper \cite{26} has the merit to present an approach to the cosmological constant problem 
very different from the current ones present in the literature based on supersymmetry, where unfortunately no supersymmetry has been 
at present day detected at LHC and thus different scenarios must be explored. In this line, it should be also noticed the recent paper
\cite{27}, where a dynamical cosmological constant is introduced, in 
a background obtained by means of an extension of the general relativity in terms of the
Ashtekar variables, where a new uncertainty relation between the dynamical $\Lambda$ and the Chern-Simons time emerges.

\section{Conclusions and final remarks}

In the usual approach to the cosmological constant problem, the vacuum energy is depicted as a summation over all possible vacuum contributions
from different energy scales present in physics (Planckian fluctuations, QCD contributions and so on.),
since thanks to the equivalence principle, all forms of energies do gravitate. Severe
cancellations and the
presence of supersymmetry is invoked in order to obtain the fine tuning necessaryy to solve the 
so called 'vacuum catastrophe'.
However, it should be noticed that when the equivalence principle is invoked, vacuum energies cannot merely summed as happens in a 
flat spacetime and the explicit expression for the vacuum energy
leading to the cosmological constant equation of state is constrained by general relativity. In fact, the natural arena for the equivalence principle is 
provided by general relativity, where the equations are not linear
and the vacuum energies modify the spacetime metric in a non-trivial way. A second important question is that in the usual treatment 
of the cosmological constant as vacuum energy it is not often clear if the vacuum contributions considered effectively satisfy
the suitable equation of state $p_{\Lambda}=-c^2{\rho}_{\Lambda}$ and not for example
the radiation one, while this fact should be carefully checked. As an example, for an harmonic oscillator, the vacuum energy looks like
$\hbar\omega/2$ in the continuum limit, and if we consider modes within a box of size $L$ we have
$\omega\sim c/L$ leading, thanks to (\ref{10}) to a radiation field equation of state rather than to the one of the cosmological
constant.
In our approach the
equation of state for $\Lambda$ is fixed from the onset by formula (\ref{10}). 
According to the line present in 
\cite{26,27}, a new way to treat the vacuum energy is urgent. In this paper, we propose a new approach, based on the results
present in \cite{3,4} in particular. The basic idea is that the cosmological constant has a 
quantum origin and can be depicted only when
Planckian fluctuations are correctly taken into account. In this regard, a statistical description of the cosmological constant 
$\overline{\Lambda}$ emerges with a dependence on the scale at which the spacetime is averaged. To this purpose, the spacetime at Planckian scale is fluctuating and we depict these fluctuations in terms of an average over a proper volume $V(L)$ composed of 
inhomogeneous metrics, by using a suitable modification of the Buchert formalism and 
without introducing a quintessence field. Another important ingredient in our modeling,
often missing in the literature, is that general relativity provides an expression for the quasi-local energy within a sphere
$L$ given by the Misner-Sharp mass $E_{ms}$. In particular, in usual thermodynamical systems within a volume $V$, there is not a priori relation between the energy $E$ and the volume $V$. Conversely, general relativity
it gives a further geometric information 
providing, by means of $E_{ms}$, a relation between $E_{ms}$ in a spherical box and the area of the box. As an example, the ADM black hole mass $M$
is related to the areal radius $R$ by the famous relation $R=2GM/c^2$. These constraints must be taken into account for a sound
description of the classical de Sitter universe: this is what we have done for example in equation (\ref{8}).
As a consequence of our setups, the classical de Sitter spacetime arises at the decoherence scale $L=L_D$, representing
an absolute minimum for our 'phenomenological' expression for $U(L)$. This absolute minimum represents the crossover to 
classicality. Moreover, since $L=L_D$ is an absolute minimum for $U(L)$, the value for $\overline{\Lambda}$ we observe is fixed 
exactly at this decoherence length-scale, expressed in terms of the two phenomenological parameters, namely
$\{L_D,\xi\}$, by equation (\ref{51}) an can be thus an explanation to the small value for 
$\overline{\Lambda}$ we observe. As a final remark, it should be noted that the mechanism depicted in the paper for the origin of
the cosmologcial constant can in principle be also applied at the primordial inflation. There, the proper dimension of the universe before the inflation has been very small, of the order of the Planck scale and thus the potential (\ref{41}) should have gained an absolute minimum
above the Planck scale, thus representing the begin of the primordial inflation in presence of other kind of matter-radiation. This can be certainly matter for next works.

\end{document}